\documentclass[aps,prb,reprint,amsmath,superscriptaddress,amssymb,showpacs,floatfix]{revtex4-1}
\usepackage[english]{babel}
\usepackage{graphicx}
\usepackage{dcolumn}
\usepackage{bm}
\usepackage[utf8]{inputenc}
\usepackage{booktabs}
\usepackage{empheq}
\usepackage{setspace}
\usepackage{multirow}
\usepackage{color}
\usepackage{indentfirst}

\newcommand{\um}[1]{\,\rm{#1}}

\newcommand{\dl}[0]{\Delta}
\newcommand{\epk}[0]{\epsilon_k}

\newcommand{\El}{\mathcal{E}}

\makeatletter
\renewcommand{\fnum@figure}{FIG. \thefigure}
\makeatother

\begin{document}

\title{Voltage-induced metal-insulator transition in a one-dimensional charge density wave}
\author{Giuliano Chiriacò}
\affiliation{Department of Physics, Columbia University, New York, NY 10027}
\author{Andrew J. Millis}
\affiliation{Department of Physics, Columbia University, New York, NY 10027}
\affiliation{Center for Computational Quantum Physics, The Flatiron Institute, New York, NY 10010}

\date{\today}

\begin{abstract}
We present a theoretical investigation of the voltage-driven metal insulator transition based on solving coupled Boltzmann  and Hartree-Fock equations to determine the insulating gap and the electron distribution in a model system -- a one dimensional charge density wave. Electric fields that are parametrically small relative to energy gaps can shift the electron distribution away from the momentum-space region where interband relaxation is efficient, leading to a highly non-equilibrium quasiparticle distribution even in the absence of Zener tunneling. The gap equation is found to have regions of multistability; a non-equilibrium analog of the free energy is constructed and used to determine which phase is preferred.
\end{abstract}

%\pacs{ 71.10.-w, 71.30.+h, 71.45.Lr}

\maketitle

\section{Introduction}

Insulator to metal transitions (IMTs) driven in thermal equilibrium by variation of temperature, strain or chemical composition  are of long-standing interest  in condensed matter physics\cite{MIT:review}. Recently, attention has shifted to non-equilibrium transitions driven by application of strong optical \cite{Aver:optCu,Aver:optLa,Aver:opt-rev,Mazur,laser:gold,Basov:review}, terahertz \cite{Cavalleri:K3C60, Aver:THz} or dc \cite{Mitra06,Maecit:7,Maecit:8,Maecit:9,Maecit:9,Maecit:10,Maecit:11,GaTaSe} electric fields.  Two broad classes of transition mechanisms have been addressed in the literature:  virtual electronic  transitions causing  changes in the Hamiltonian (``Floquet engineering") and real electronic transitions, changing the electron distribution function. Typically, important effects occur when the non-equilibrium drive is comparable to some important energetic or lattice scale; for example, when Hamiltonian parameters are changed enough to drive a system through a $T=0$ phase transition, or a large enough number of valence band carriers are excited over the gap, or atomic positions are displaced by a significant fraction of the lattice constant.

In an interesting recent experiment,  Maeno and collaborators \cite{Maeno:mainp, Fuku:CRO} reported that in Ca$_{2}$RuO$_4$ \cite{Pavarini:CRO,Friedt:CROPT}, modest electric fields $F^\star\sim 40$ V/cm can suppress the metal-insulator transition temperature from $T_{IMT}=356\um{K}$  to substantially below room temperature. One might expect that the main effect of an applied dc field would be to enable carriers to tunnel across a band gap, and that the  critical electric field required to drive an IMT would have to be strong enough to produce a large number of real excitations, i.e. to  be of the order of the energy gap divided by some suitable atomic-scale length. For example, a  non-equilibrium dynamical mean field analysis of a current-driven Mott insulator \cite{Amaricci12} found important effects  when applied fields $F$ were large enough that the voltage drop across one unit cell $\sim eFa$ was comparable to the Mott gap.

From this point of view, the value of the critical field required to drive the transition in Ca$_2$RuO$_4$ is remarkably small: the electronic energy gap of the insulator is $\Delta \approx 0.2-0.6\um{eV}$  so the experimentally applied field $\sim 40$ V/cm corresponds to a length $L^\star=\Delta/F^\star\sim 10^5$ lattice constants. Landau-Zener tunnelling \cite{ZL:Landau, Zener:two} leads to an excitation rate proportional to $e^{-\Delta^2/WeFa}$ where $W$ is a measure of the bandwidth, and thus to a field scale $\Delta^2/Wea$ which is parametrically smaller (in the limit of small gap) but still set by fundamental atomic-scale energies. In the Ca$_2$RuO$_4$ case inserting $W\sim 1.5\um{eV}$ into the Zener formula would yield a length $L^\star\sim 10^4$ lattice constants. The results of Refs. ~\cite{Maeno:mainp, Fuku:CRO} therefore motivate further investigation into alternative mechanisms for non-equilibrium metal-insulator transitions.

In this paper we analyse a mechanism by which an applied electric field can change an electronic distribution function without directly exciting carriers over a gap. The key point is that interband relaxation is strongly dependent on position in momentum space, so that an electric field can shift carriers away from points of rapid relaxation, leading to a population imbalance that is set by comparing the electric field to a relaxation time, rather than an energetic scale. The resulting effects are power law, not exponentially small, in the field strength. To investigate this issue we use a Boltzmann equation plus mean field analysis of a one dimensional model of spinless fermions with a charge density wave instability \cite{CDW:Peierls}. The field-induced renormalization of the critical temperature can be large, eventually pushing the linear instability to density wave order down to zero temperature. However the destabilization of the density wave state is weaker, leading to a bistable behavior, characterized by the coexistence of both metallic and gapped stable phases. We emphasize that our work is not intended to specifically model the experiments of Refs. \cite{Maeno:mainp, Fuku:CRO}; rather it is a theoretical study of an alternative mechanism, motivated by the key features of the experiments of Maeno \emph{et al}. Our work is complementary to recent work \cite{Han} studying the IMT  when the Zener tunneling is important.

The rest of this paper is organized as follows: in section \ref{model} we present the model we study, analyze the scattering mechanisms and write the Boltzmann equation, which we solve in section \ref{solutions}; in section \ref{results} we report the results for the gap and in section \ref{stability} we study the stability of the phases. Section \ref{Conclusion} provides a summary, conclusions, and prospects for future work. Appendices provide technical details.

\section{Model\label{model}}

\subsection{Hamiltonian and kinetic equation}

We study a model of a single band of spinless fermions moving on a one-dimensional lattice of lattice constant $a$ with energy dispersion $\varepsilon_k$; we assume the band is half filled and that the fermions are subject to an interaction parameterized by the coupling constant $G$ that leads to a commensurate (period $\pi/a$) site-centered charge density wave  of amplitude $\propto\dl$.

We define the electron annihilation operator on site $j$ as $c_j$, and write the mean field hamiltonian in the Fourier basis appropriate to the doubled unit cell as:
\begin{equation}
H=\sum_k\begin{pmatrix}c^\dagger_k & c^\dagger_{k+Q}\end{pmatrix}
\begin{pmatrix}\varepsilon_k & \dl \\ \dl & \varepsilon_{k+Q}\end{pmatrix}
\begin{pmatrix}c_k \\c_{k+Q}\end{pmatrix},
\label{HMF}
\end{equation}
where $c_k=\frac1{\sqrt{N}}\sum_je^{ikj}c_j$, the wavevector $k$ is measured in units of $1/a$ and $Q=\pi$.

The eigenstates of the Hamiltonian are conduction ($c$) and valence ($v$) bands with energies
\begin{equation}\label{bandE}
E_k^{c/v}=\frac{\varepsilon_k+\varepsilon_{k+Q}}{2}\pm
\sqrt{\left(\frac{\varepsilon_k-\varepsilon_{k+Q}}{2}\right)^2+\Delta^2}.
\end{equation}
The minimum band gap is  $2\dl$ and we choose $\varepsilon_k$ such that the point of minimum gap is   $k=\pm Q/2$.

The mean field equation for the gap is
\begin{equation}\label{GapE0}
1=\frac{G}{N}\sum_k\frac{n^v_k-n^c_k}{\sqrt{\left(\frac{\varepsilon_k-\varepsilon_{k+Q}}{2}\right)^2+\Delta^2}},
\end{equation}
where $n^{v/c}_k$ are the occupations of states $k$ in the valence/conduction band.

In equilibrium at $T=0$, $n^v=1$, $n^c=0$ and perfect nesting of one-dimensional band structures means that at $k=Q/2=\pi/2$,  $\varepsilon_k=\varepsilon_{k+Q}$;  thus the logarithmic divergence of the sum in Eq. \eqref{GapE0} at $\dl\rightarrow0$ implies the existence of a solution with $\dl\neq0$. As $T$ is increased, $n^v$ decreases and $n^c$ increases, eventually  leading (within mean field theory) to a second order transition at a temperature $T_C$ set by $G$. Because  this is a one dimensional system, beyond mean-field effects will convert the transition to a crossover between a high-$T$ short ranged correlated state and  a low $T$ state described by an exponentially large, although finite, correlation length. This physics is not relevant to the considerations of this paper.

\begin{figure}[t]
\centering
\includegraphics[width=0.9\columnwidth]{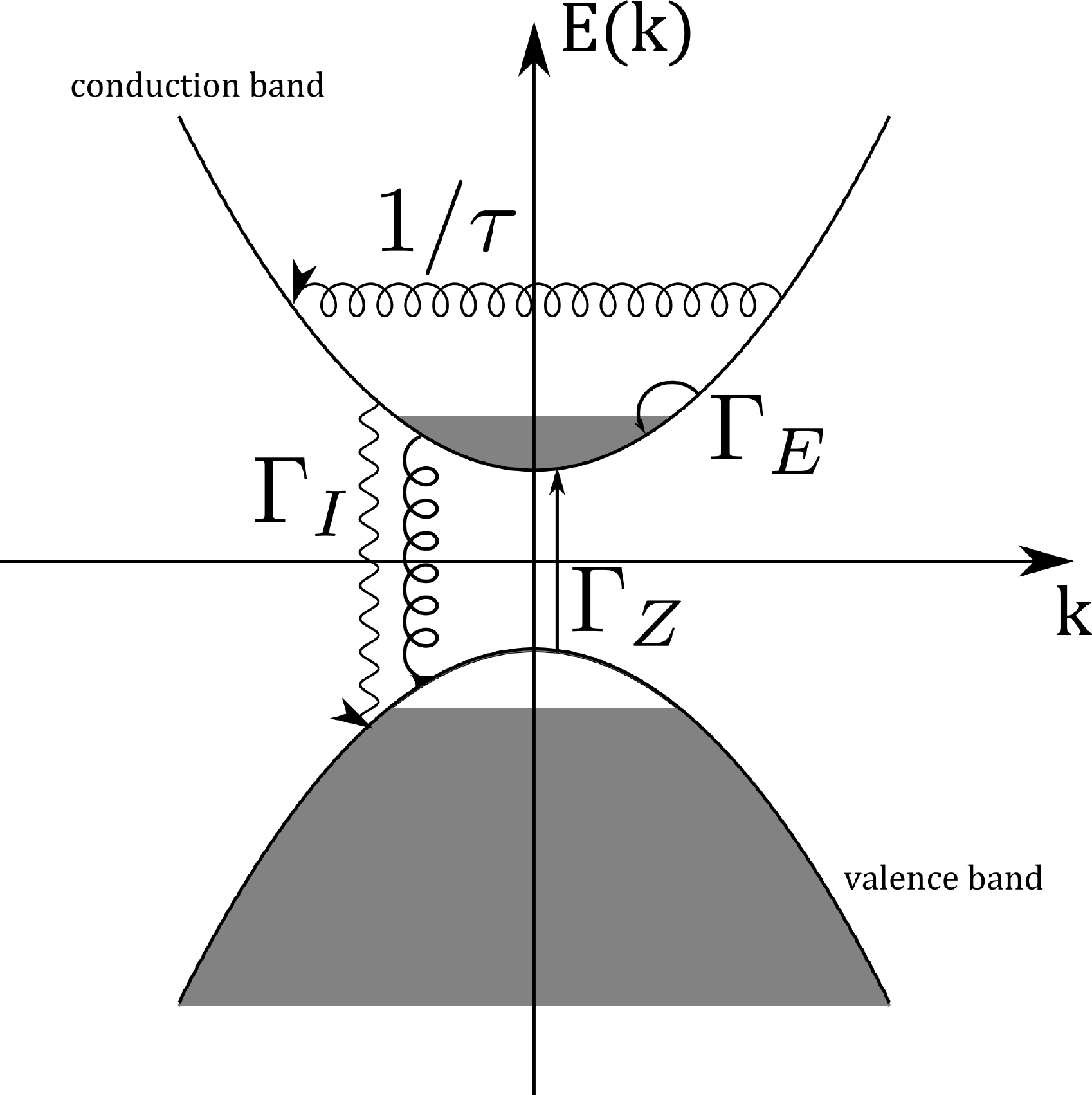}
\caption{\scriptsize{Sketch of scattering processes near the gap. The rates are: $\Gamma_I$ for the interband scattering mediated by photons (wavy curves) or phonons (curly curves), $\Gamma_Z$ for the Landau-Zener tunneling, $\tau^{-1}$ for the momentum relaxation and $\Gamma_E$ for the energy relaxation.}}
\label{compscatt}
\end{figure}
We now consider how an applied electric field changes the distribution functions and thus the solution of the gap equation. To this end we write and solve Boltzmann transport equations for the steady state conduction and valence band occupation $n^{c/v}_k$. The transient state, while interesting in its own right, is beyond the scope of this paper. The crucial ingredients of a Boltzmann equation are the acceleration of the carriers by the applied field $F$, a momentum relaxation process (which we consider to come from energy-conserving scattering with rate $\tau^{-1}$), an interband scattering that changes the number of particles in each band (rate $\Gamma_I$), the Landau-Zener tunneling (rate $\Gamma_Z$) and an intraband energy relaxation process (rate $\Gamma_E$). Not notating the dependences of the $\Gamma$ on the distribution functions, we have
\begin{eqnarray}
\partial_tn^c_k&=&-\frac{eFa}{\hbar}\partial_kn^c_k-\Gamma_I+\Gamma_Z-\frac{n^c_k-n^c_{-k}}{\tau_k}+\Gamma^c_E;
\label{condband}\\
\partial_tn^v_k&=&-\frac{eFa}{\hbar}\partial_kn^v_k+\Gamma_I-\Gamma_Z-\frac{n^v_k-n^v_{-k}}{\tau_k}+\Gamma^v_E.
\label{valenceband}
\end{eqnarray}

The Boltzmann equations \eqref{condband}--\eqref{valenceband} are coupled nonlinear equations and the general solution is complicated. To simplify the presentation without losing essential features we assume particle-hole symmetry in the electron dispersion and scattering amplitudes. In this case, $n^v=1-n^c$ and the two equations can be collapsed to one. For notational simplicity we choose the origin of $k$ to coincide with the gap minimum, assume $\varepsilon_k=-2t\sin~k$, define the Fermi velocity $v_F=2t$ and normalize all energy scales to the hopping term $t$.

We exploit the symmetry under $k\leftrightarrow -k$ to separate the odd and even parity parts of the distribution, defining
\begin{equation}
n_{e/o}=\frac12(n^c_k\pm n^c_{-k}),
\label{nevodddef}
\end{equation}
and rearrange  the equations  to make the physically interesting limit $\Gamma\tau\ll1$ more transparent, obtaining%
\begin{subequations}\label{BEoe}
\begin{gather}\label{BEo}
\mathcal{E}\partial_kn_e=-n_o-\gamma_{I,o}+\gamma_{Z,o}+\gamma_{E,o};\\
\label{BEe0}
\mathcal{E}\partial_kn_o=-\gamma_{I,e}+\gamma_{Z,e}+\gamma_{E,e}.
\end{gather}
\end{subequations}
Here we have neglected the dependence on the momentum of the elastic scattering rate, defined the dimensionless electric field $\El=eFa\tau/(2\hbar)$, and normalized the even/odd part of the $i$th scattering rate $\gamma_{i,e/o}\equiv\Gamma_{i,e/o}\tau/2$.

If all of the $\Gamma\ll\frac{1}{\tau}$ and the electric field is not too large, Eq. \eqref{BEo} implies $n_o=-\mathcal{E}\partial_kn_e$ and Eq. \eqref{BEe0} becomes
\begin{equation}
\label{BEe}
\mathcal{E}^2\partial_k^2n_e=\gamma_{I,e}-\gamma_{Z,e}-\gamma_{E,e}.
\end{equation}

\subsection{Scattering processes}

In this subsection we specify important features of the  inelastic scattering processes (sketched in Fig.~\ref{compscatt}). Details are provided in Appendix A.

We take the interband process $\Gamma_I$ to arise from scattering involving a bosonic mode (photon or optical phonon) and calculate it using the standard Fermi golden rule
\begin{equation}\label{gI}
\Gamma_I(k)=\mathcal{A}_k^2\left[n^c_k\left(1-n^v_k\right)\left(1+b\right)-n^v_k\left(1-n^c_k\right)b\right].
\end{equation}
Here $b$ is the Bose distribution at energy $\Delta E_k=E_k^c-E_k^v\equiv2E_k$ and $\mathcal{A}_k$ is the transition matrix element. We have assumed that the interband scattering is essentially vertical (momentum conserving); this is clearly justified in the case of optical emission and is  a reasonable approximation for optical phonons when the phonon energy $\omega_{ph}$ is much smaller than the bandwidth so that the process is only important for electrons in a range $\delta k\sim\omega_{ph}/v_F\ll1$ of the gap minimum. We recast Eq. \eqref{gI} using the definitions and approximations of section II.A
\begin{equation}\label{gIe}
\gamma_{I,e}=\gamma_b(n_e^2+2n_eb-b);\qquad \gamma_b\equiv \mathcal{A}_k^2\tau/2.
\end{equation}
This form will be used in our subsequent analysis.

In Eq. \eqref{gIe} the matrix element $\mathcal{A}_k$ %, which is  hidden in the coefficient $\gamma_b$,
plays a crucial role. On physical grounds we expect $\mathcal A_k$ to drop rapidly as $k$ is shifted away from the gap: for optical emission the probability is $\mathcal{A}_k^2\sim \Delta^2/E_k$, while when the conduction-valence band energy difference becomes greater than $\omega_{\textrm{ph}}$, the multiple phonon emissions required for down scattering lead to a rapid suppression: in other words, interband relaxation is only efficient for carriers with energies near the conduction band minimum (valence band maximum). This is  important because  in equilibrium the ``up scattering" [second term in Eq. \eqref{gI}] and  ``down scattering" (first term) processes cancel, as can be verified by substituting the appropriate  distribution functions in Eq. \eqref{gI}. At low $T$, up scattering is controlled by the probability of finding a thermally excited boson of the correct energy while down scattering is constrained by fermion occupancies. Out of equilibrium the field $F$ sweeps carriers away from the conduction band minimum/valence band maximum (gap) into regions where the interband relaxation is less efficient, leading to changes in population even without Zener tunneling.

We now study the energy relaxation term $\Gamma_E$. We imagine that the system is in contact with a reservoir held at temperature $T$ with which it can exchange energy in very small increments $\delta E$. This leads to an intraband scattering mechanism whose rate is evaluated with the Fermi golden rule (see Appendix A for more details):
\begin{equation}
\label{gEe}
\gamma_{E,e}=\gamma_RT\partial^2_En_{e}+\gamma_R\partial_E \left(n_{e}-n_{e}^2\right),
\end{equation}
where $\gamma_R$ is a dimensionless rate (which includes the exchanged energy $\delta E$ normalized to $t$) and $\partial_E$ is the derivative with respect to the conduction band energy. Notice that $\gamma_{E,e}$ vanishes if $n$ is the Fermi-Dirac distribution.

Finally we briefly address Landau-Zener tunneling: it promotes electrons from valence to conduction band with a rate proportional to $e^{-\pi\dl^2/(2teFa)}$; thus it is exponentially small in $\dl$ and for the values of the electric field considered here it is relevant only for $\dl\lesssim0.005$.

\subsection{Final form of kinetic equation}

Substituting the expressions for the interband and energy relaxation into Eq. \eqref{BEe}, neglecting the Zener tunneling term and introducing $v\equiv\partial_kE_k$, we have
\begin{eqnarray}
&&\gamma_b\left(n_{e}^2+2n_{e}b-b\right)=\left(\mathcal{E}^2v^2+\gamma_RT\right)\partial_E^2n_{e}+\nonumber \\
&&+(\gamma_R(1-2n_e)+\mathcal{E}^2v\partial_Ev)\partial_En_{e}.
\label{finaleq}
\end{eqnarray}
In equilibrium, the left hand side of Eq. \eqref{finaleq} vanishes if $n$ is the Fermi-Dirac distribution $n_{\textrm{FD}}$, as a detailed balance requires. Equation \eqref{finaleq} is the basis for our subsequent analysis.

\section{Boltzmann equation analysis \label{solutions}}

Even though Eq. \eqref{finaleq} cannot be solved analytically, progress can be made in particular limits:

\textit{Zero gap case}. Let us first assume $\Delta=0$ so there is no charge density wave (CDW) order. In this case the interband scattering is irrelevant, and for energies near the Fermi level $v=v_F$, so that we have
\begin{equation}
\left(v_F^2\mathcal{E}^2+\gamma_RT\right)\partial^2_En_{e}+\gamma_R\partial_E n_{e}\left(1-2n_{e}\right)\approx0.
\label{nogap}
\end{equation}
The solution is a thermal distribution with an effective temperature  $T_{\textrm{eff}}$ given by
\begin{equation}
T_{\textrm{eff}}=T+v_F^2\mathcal{E}^2/\gamma_R,
\label{Teff}
\end{equation}
reflecting the balance between Joule heating of the electrons ($v_F^2\mathcal{E}^2$) and energy dissipation into the reservoir ($\gamma_R$). This Joule heating leads to a suppression of the linear instability to CDW order, which now occurs at the reduced value
\begin{equation}
\label{TCE}
T_C(\El)=T_C(\El=0)-v_F^2\El^2/\gamma_R.
\end{equation}
$T_C$ is suppressed to 0 when $\El=\El^{\star}_T\equiv\sqrt{\gamma_RT_C}/v_F$.

\begin{figure}[t]
\centering
\includegraphics[width=0.96\columnwidth]{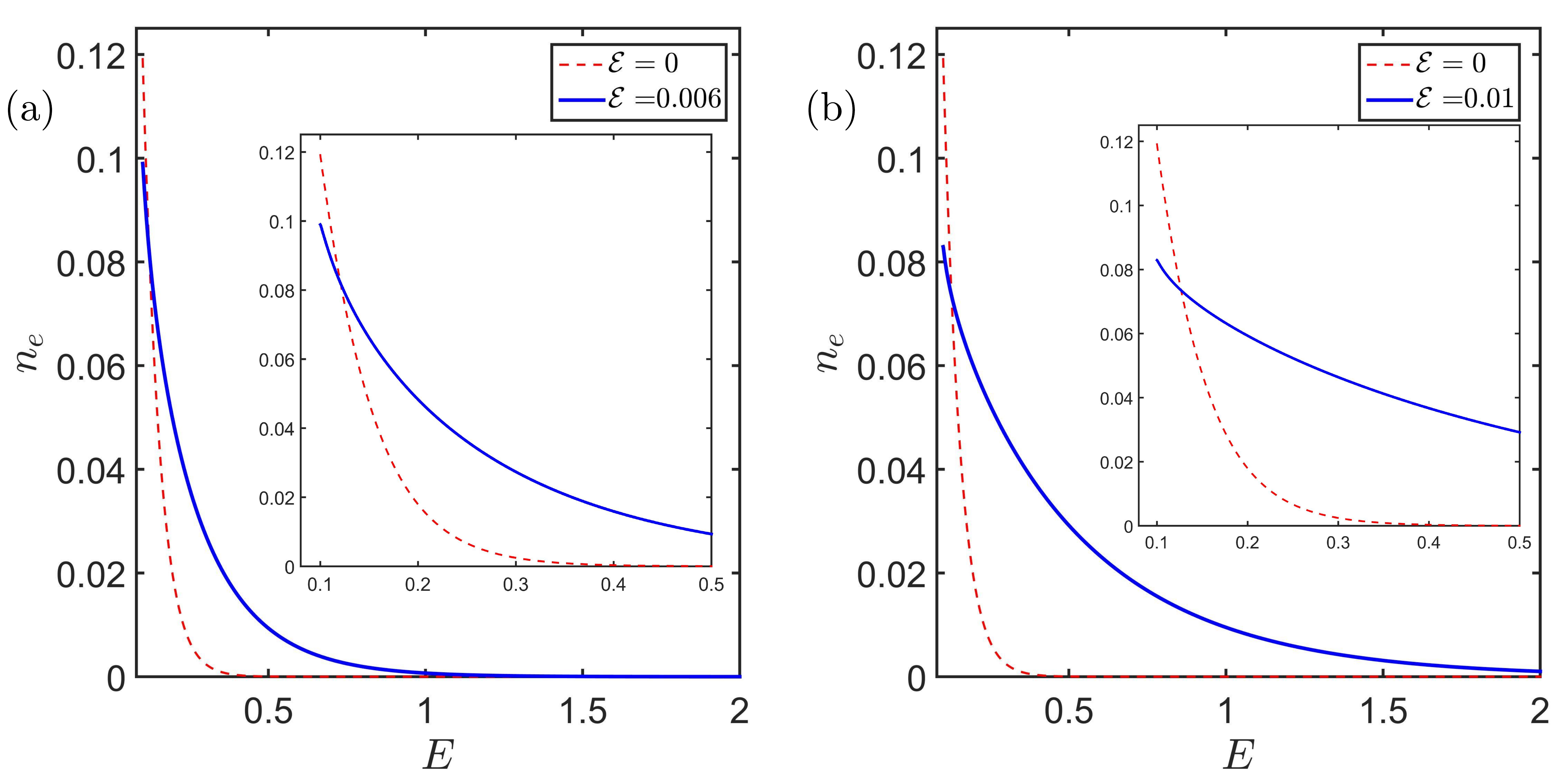}
\caption{\scriptsize{(Color online) Even part of the distribution $n_e$ as function of the energy $E$ for $\dl=0.1$, $T=0.05$, $\gamma_R=0.001$, $\omega_{ph}=0.1$ and values of the normalized electric field corresponding to $\El=0.006$ (a) and $\El=0.01$ (b). The insets show $n_e$ in a range of energies closer to the gap.}}
\label{nk}
\end{figure}
\textit{$\dl>T$ case}. For $\Delta \neq 0$, Eq. \eqref{finaleq} has an interesting structure: its right hand side conserves the particle number in the conduction band while its left hand side (interband transitions) does not; this means that the steady state solution must be such that the average over energies of the left hand side of Eq. \eqref{finaleq} vanishes.  When $\Delta>T$, $\gamma_Z$ is negligible and we expect the contribution from $\gamma_b$ to be small and vanish rapidly for  $E\gtrsim \Delta$. Thus we set the left hand side of Eq. \eqref{finaleq} to zero and neglect the quadratic terms, finding
\begin{equation}\label{nogb}
\partial^2_En_e=-\frac{1+\mathcal{E}^2v\partial_Ev/\gamma_R}{T+\mathcal{E}^2v^2/\gamma_R}\partial_En_e.
\end{equation}

As shown in detail in Appendix B, Eq. \eqref{nogb} determines $n_e$ up to a multiplicative constant, which can be found by requiring that the upscattering and downscattering  terms in Eq. \eqref{finaleq} balance:
\begin{equation}\label{balance}
\sum_k(n_e^2+2n_eb-b)\gamma_b=0.
\end{equation}
In the low $T$ limit the integrals are confined to $E\sim \Delta$; in this region $n_e\sim b^{1/2}\sim e^{-\dl/T}$ . The consequence is that $n_e$ (see Fig. 2) is of the order of $e^{-\dl/T}$ multiplied by a factor depending on energy, temperature and field
\begin{equation}\label{neasy}
n_e\sim e^{-\frac{\dl}T}f(E,T,\El).
\end{equation}
An inspection of Eq. \eqref{nogb} at $E\gg\dl$ shows $f\sim e^{-(E-\dl)/T_{eff}}$ so that the distribution function is pseudothermal, spread over a wider energy range $\sim T_{eff}$ and comprises more particles than in equilibrium; however at $E-\dl\lesssim\dl$, $f\lesssim1$, implying that for small energies $n_e$ is less than its equilibrium value. Because the gap equation \eqref{GapE0} weights more low energies, the increase in $n_e$ (relative to equilibrium) at high energies can be compensated by the decrease at low energies leading to a small net reduction of the gap, in particular at low $\dl$. This follows from the weak coupling model considered here; the gap decrease would be larger if the CDW were not driven by a low energy instability.

Moreover, Eq. \eqref{neasy} implies that any modification to the gap equation \eqref{GapE0} will be exponentially small in $\dl/T$, making an insulating phase $\dl\gg T$ hard to destabilize even at high $\El$; on the other hand, the instability of the metal phase is suppressed to very low temperatures for $\El^2v_F^2/\gamma_R\sim T_C$. This leads to a bistability region in a range of $T$ and $\El$, characterized by a coexistence of insulating and metal phase.

\textit{$\dl<T$ case}. When the gap is non-zero but smaller than the temperature, the terms that are quadratic in $n_e$ cannot be neglected in principle and no analytic solution is obtainable. However we expect the occupation in this regime to be a crossover between the Fermi-Dirac distribution with effective temperature $T_{eff}$ and the solution given by Eq. \eqref{neasy}.

\section{Numerical Results}\label{results}

\begin{figure}[t]
\hspace*{-0.03\textwidth}
\centering
\includegraphics[width=0.94\columnwidth]{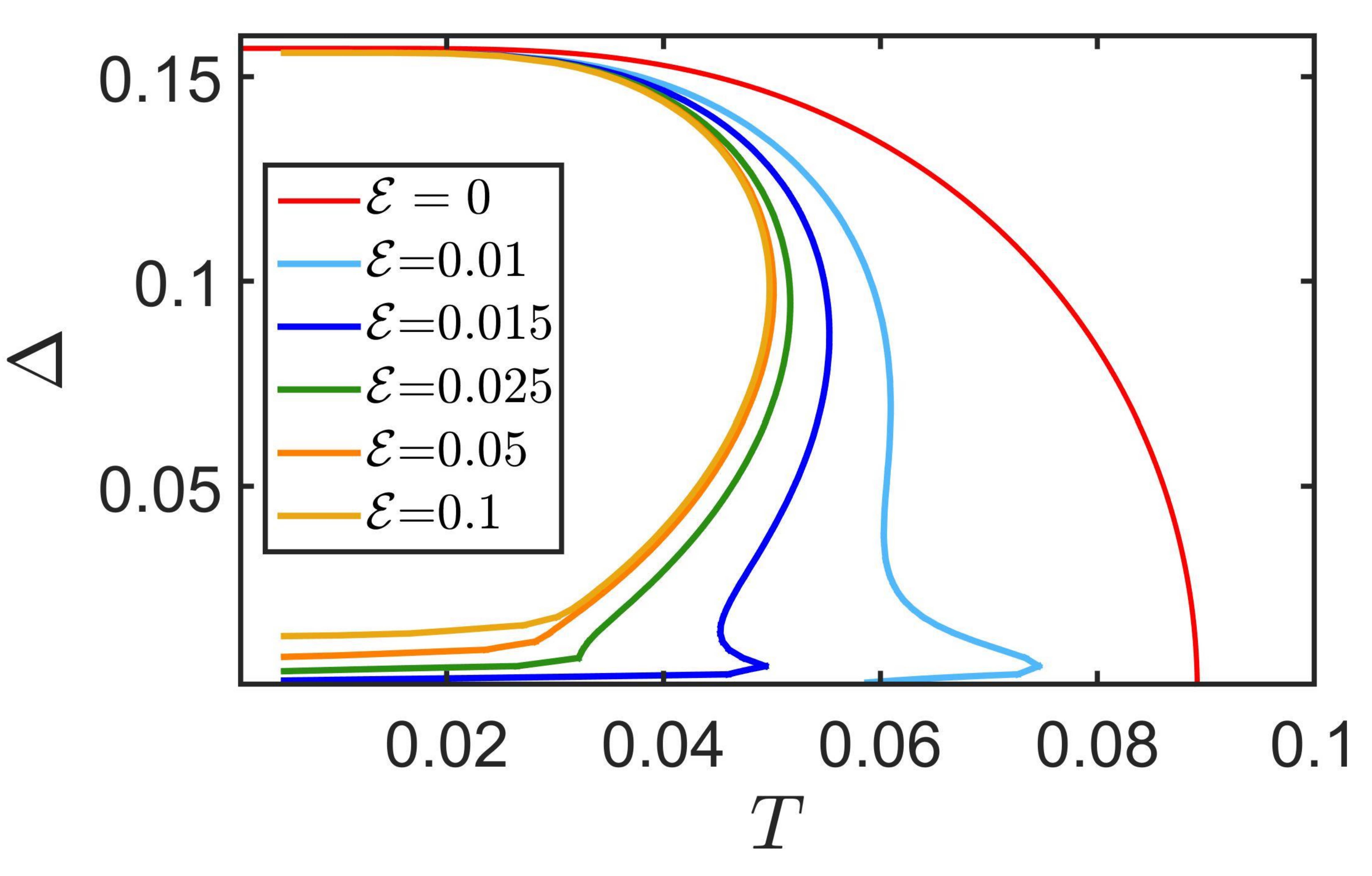}
\caption{\scriptsize{(Color online) Plot of the gap $\dl$ as function of the temperature $T$ for several values of the electric field $\El$ and for $G=1.6$ (corresponding to $T_C\approx0.09$), $\omega_{ph}=0.1$, $\gamma_R=0.001$. The values of the IMT transition temperature are $T_1\approx0.055$ for $\El=0.015$ and $T_1\approx0.049$ for very high $\El$.}}
\label{DTgap}
\end{figure}

We solve numerically Eq. \eqref{finaleq} in the general case, choosing  reasonable values for the parameters $\omega_{ph}/t\sim0.1$ and $\gamma_R$, $\gamma_b$ ($\sim10^{-3}$, $10^{-4}$). We then substitute $n_e$ into Eq. \eqref{GapE0} and self-consistently solve for $\dl(T)$, which is plotted for different $\El$ (Fig. \ref{DTgap}). When $\El^2/\gamma_R\gtrsim1$ no appreciable change occurs by increasing the electric field, consistently with the run-away heating regime appearing when $T_{eff}$ is of the order of the bandwidth.

We observe the bistability predicted in section ~\ref{solutions}: for a given $\El$ there exists a range of temperatures $T_C(\El)<T<T_1(\El)$ for which a stable high $\dl$ insulating phase coexists with a stable metal $\dl=0$ phase. Notice that the gapped phase and the metal phase are not analytically connected through a stable phase: thus any switching between the two phases occurs with a jump in $\dl$, which corresponds to a first order phase transition. The value of the insulator to metal transition temperature $T_1(\El)$ decreases as $\El$ increases, but is limited from below by $T_{1\infty}=T_1(\El\rightarrow\infty)$: even at high $\El\sim0.1$ the insulating state survives for $T<T_{1\infty}$, see Fig. \ref{DTgap}.

At low $\El$, we also observe the appearance of a stable low $\dl$ phase (which is not insulating, being characterized by $\dl\ll T$) for $T_2(\El)<T<T_3(\El)$; $T_3(\El)$ can be larger or smaller than $T_1(\El)$ and the difference $T_3(\El)-T_2(\El)$ decreases with $\El$ until it vanishes. This phase is caused by the previously mentioned effect that a $n_e(E\sim\dl)<n_{FD}$ has on the gap equation at low $\dl$.

\section{Stability Analysis \label{stability}}

To study the stability of the different phases, we multiply Eq. \eqref{GapE0} by $\dl$ obtaining
\begin{equation}
\label{Fgap}
\dl-\dl\frac GN\sum_k\frac{n_k^v-n_k^c}{E_k}\equiv\dl-\Phi[\dl,T,\El]=0.
\end{equation}
We interpret the left hand side of Eq. \eqref{Fgap} as the derivative with respect to $\Delta$ of a nonequilibrium ``free energy" and integrate it (in practice the integral is performed numerically) obtaining
\begin{equation}
\label{Fener}\mathcal F[\dl,T,\El]=\int_0^{\dl} d\dl'(\dl'-\Phi[\dl',T,\El]).
\end{equation}

The stationary points of $\mathcal F$ solve the gap equation: the minima correspond to stable solutions and the maxima to unstable ones.

In Fig. 4(a )we plot $\mathcal F$ for three values of $T$ both in equilibrium and out of equilibrium. At $T<T_C$ and $\El=0$ we observe the usual behavior of the equilibrium free energy of a system below its critical temperature; by increasing $\El$, the metal phase becomes locally stable, a local maximum appears at intermediate values of $\dl$ (corresponding to the unstable middle branch of Fig.~\ref{DTgap}) and the free energy of the high $\dl$ phase increases. For $T<T^{\star}(\El)$ the insulating phase is ``energetically favored'' compared to the metal phase, so it is globally stable; for $T^{\star}(\El)<T<T_1(\El)$ the metal phase is ``favored'' and the insulating phase becomes only locally stable, eventually disappearing at $T>T_1(\El)$; the temperature $T^{\star}(\El)$ is defined by $\mathcal F[\dl(T^{\star}),T^{\star},\El]=0$.

\begin{figure}[t]
\centering
\includegraphics[width=0.94\columnwidth]{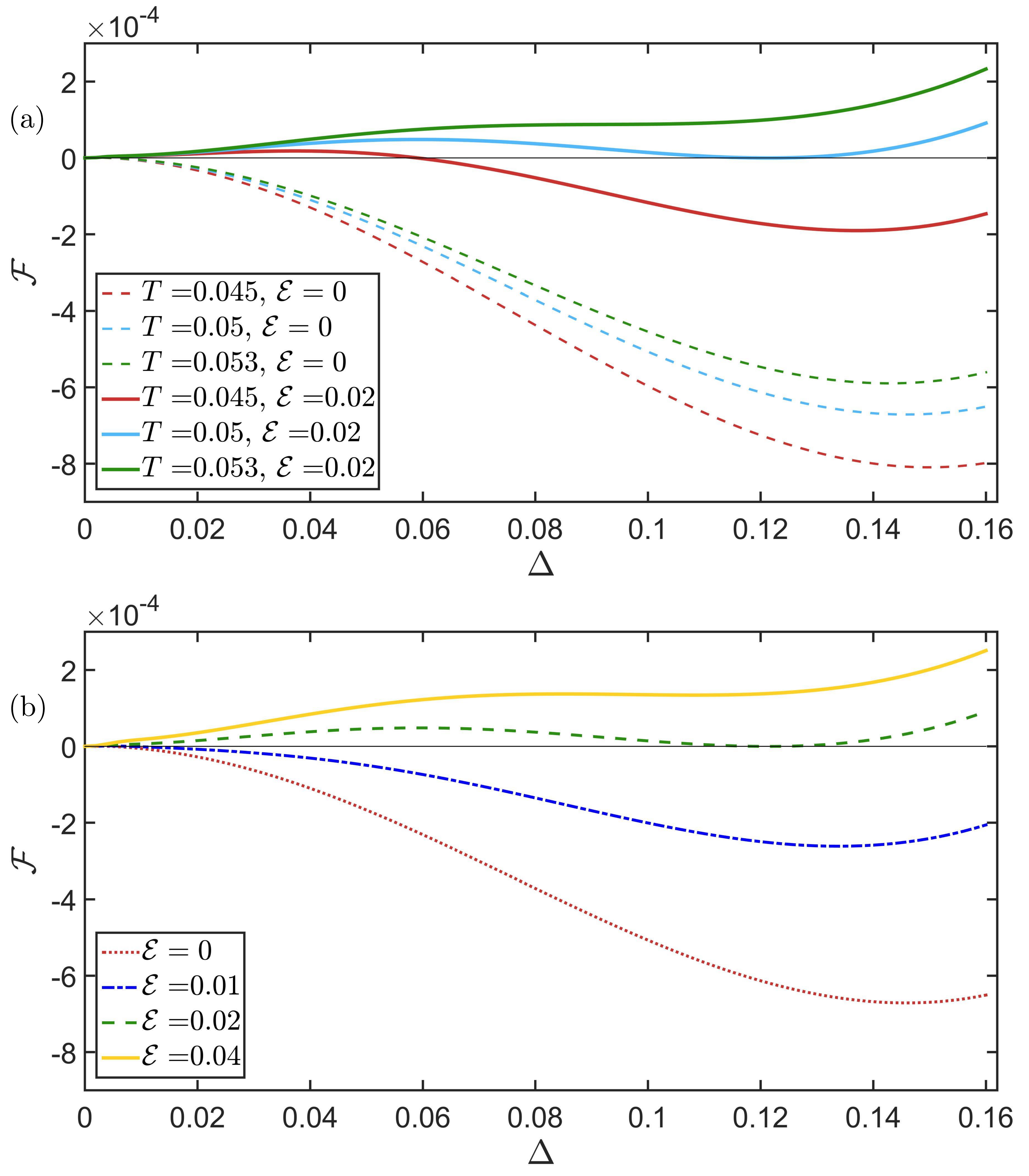}
\caption{\scriptsize{(Color online) Free energy $\mathcal F$ as function of the gap $\dl$ for $G=1.6$, $\omega_{ph}=0.1$ at electric field $\El=0.02$ (a) or temperature $T=0.05$ (b). In (a), the IMT temperature is $T_1\approx0.053$ at $\El=0.02$, while $T^{\star}\approx0.05$. In (b), $T_1(\El=0.04)\approx0.05$ and $T^{\star}(\El=0.02)\approx0.05$.}}
\label{LFE}
\end{figure}
We observe that this  energy functional  implies that there may be hysteresis when the system is tuned through the transition. For example, consider the  system to be initially in the insulating phase below $T^{\star}(\El)$, with $\El$ strong enough to exclude a stable low $\dl$ phase. On heating, the insulating phase becomes metastable for $T^{\star}(\El)<T<T_1(\El)$ and a sufficiently strong perturbation can make the system switch to the metal phase; if no perturbation occurs, the phase transition occurs at $T=T_1(\El)$. If the system is now cooled down, it remains in the metal phase down to $T=T^{\star}(\El)$ and below this temperature the metallic state is metastable and the system could switch back to the insulating phase under a suitable perturbation. A similar hysteresis cycle occurs at fixed $T$ by varying $\El$ [Fig. 4(b) for the free energy]. A detailed study of the dynamics of the phase switching requires an analysis of nucleation processes which is  beyond the scope of this paper.

\section{Summary and Conclusions\label{Conclusion}}
In summary, we have shown that a nonequilibrium drive may change the distribution function of a correlated insulator by sweeping carriers from regions of rapid interband relaxation to regions where the relaxation is less efficient. The ratio between electric field strength and a suitable relaxation rate affects the properties of the resulting distribution, which has much more weight at high energies, but less at the low energies that dominate the gap equation, and is still exponentially small in $\dl/T$; despite the parametrically large change in distribution function, the gap magnitude is only weakly affected by the field at $T\ll\dl$. Therefore the electric field is less effective in destabilizing the gap than Joule heating is in stabilizing the metallic phase, leading to a region of bistability, in which both the zero gap and large gap phases are locally stable, and thus to a first order transition in the presence of the field.

A key finding is hysteresis in the behavior when viewed as a function of  electric field strength. The hysteresis we predict should be observable in simple two-terminal experiments and indeed hysteretic behaviors in the current-voltage curve have been reported \cite{Maecit:8,Maeno:mainp}. Further, the  broad pseudothermal conduction band distribution we predict mays be observable in photoemission experiments conducted under conditions of current flow.

We remark here on the relation of our results to those obtained by Han and coworkers \cite{Han} on essentially the same model, but in a different and complementary limit. The two key differences are that Han et. al. consider energy relaxation arising from a fermionic bath, whereas in our work the energy relaxation is provided by a bosonic bath (acoustic phonons). Also we focus on field-induced changes in the electronic distribution function; these effects were not considered in Ref. [\onlinecite{Han}] and are relevant at much lower fields than the Zener tunneling on which Han \emph{et al.} focus. It is also important to note that our results  depend on the presence of thermally excited carriers. Our finding  that the CDW phase is always stable at $T=0$ arises from our  neglect of Zener tunneling.

The results presented here were motivated by the experiments on Ca$_2$RuO$_4$, but the physics we find may not be operative in Ca$_2$RuO$_4$. In the model studied here, the stabilization of the metallic phase is due to Joule heating of the electrons and the metastability  arises because the insulating phase is affected less by the field than the metallic phase. While the electron temperature has not been directly measured in Ca$_2$RuO$_4$, Joule heating of the entire sample was found not to be significant and the experimental consensus is that the involved physics is not a heating effect. However,   many other materials \cite{Maecit:8,Maecit:9,Maecit:10,GaTaSe} exhibit  voltage-driven metal-insulator transitions with threshold fields that lie in the range $\sim0.3-4\um{kV/cm}$. In these materials the effect we find would be much bigger than in Ca$_2$RuO$_4$ and could play a substantial role in driving the transition.

\section*{Acknowledgements}

This work was supported by the Basic Energy Sciences Division of the United States Department of Energy under Grant No. DE-SC0012375.

\appendix

\section{Scattering processes details}

\begin{figure}[t]
\centering
\includegraphics[width=0.44\textwidth]{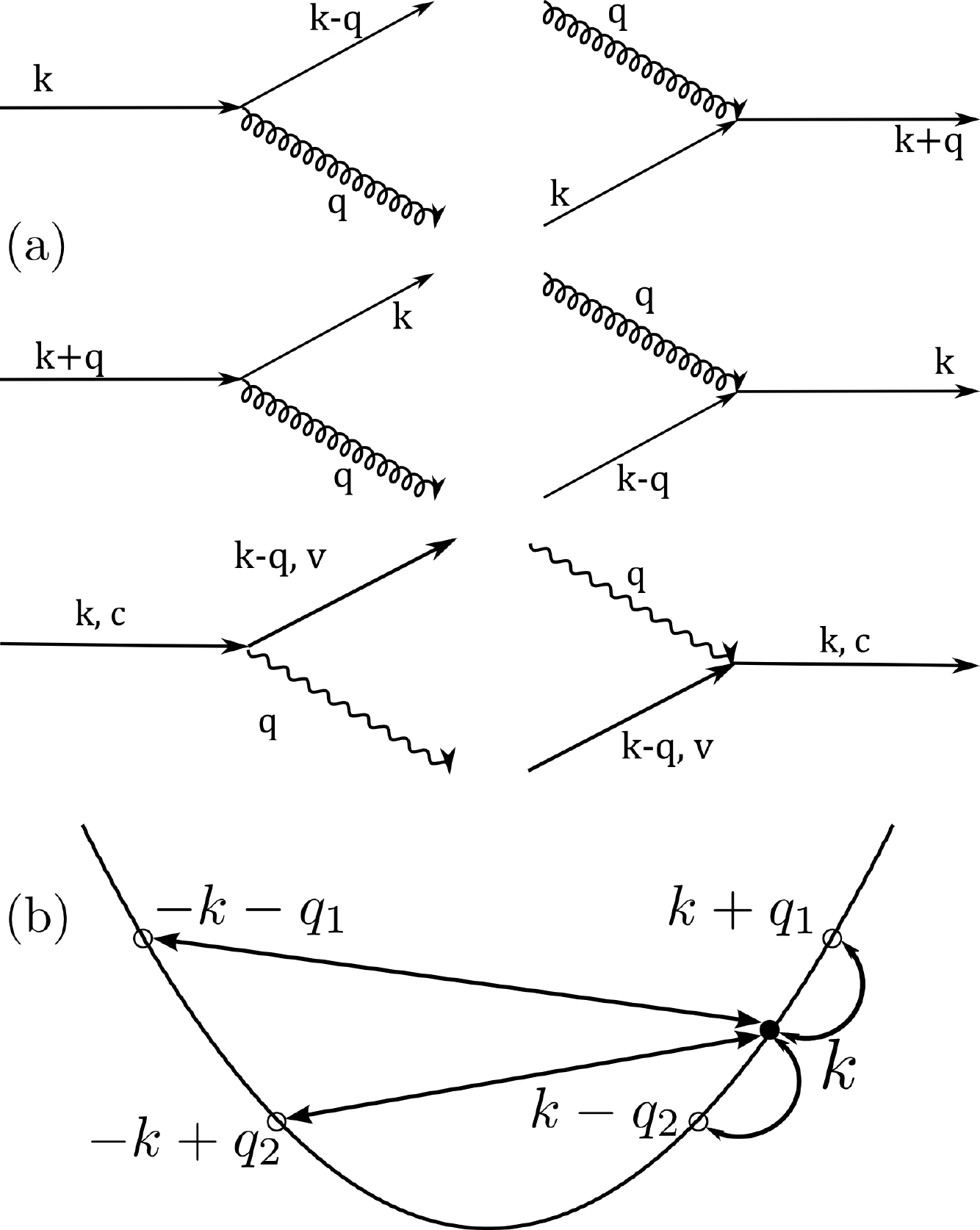}
\caption{\scriptsize{(a) Feynman diagrams of intraband scattering (top two rows) and interband scattering (third row). Solid lines represent electrons, wavy or curly lines represent the mediating bosons; the momentum is indicated for each particle, $c$ and $v$ refer to conduction and valence band. (b) Possible scattering processes for the energy relaxation mechanism $\Gamma_E$; notice the backscattering processes.}}
\label{scatmech}
\end{figure}

Here we analyze more in detail the scattering processes, depicted in Fig. \ref{scatmech}. We use standard Fermi golden rule methods to calculate the scattering rates, assuming reasonable values for the interaction strengths.

\emph{Elastic scattering.}  We imagine that this scattering arises from  interaction with acoustic phonons, which have a linear dispersion $\omega_{ac}(k)=v_s|k|$ (very similar results would be obtained if the scattering came from randomly positioned weak impurities). The sound velocity is typically low $v_s\ll v_F$ so that the phonon energy is negligible compared to the typical electron energy and thus the scattering conserves energy and occurs between states with opposite momentum. The matrix element of the transition contributes $\tau_b\propto\tau_0\partial_k\epk$ to the scattering time ($\tau_0\sim10^{-12}\,\rm{s}$); the change rate of conduction/valence electron is
\begin{gather}
\Gamma^{c/v}_{el,k}=\frac{(1-n^{c/v}_k)n^{c/v}_{-k}[1+2b(2v_s|k|)]-(k\leftrightarrow-k)}{\tau_b}\notag \\
=-\frac{n^{c/v}_k-n^{c/v}_{-k}}{\tau_b\tanh(v_s|k|/T)} \equiv -\frac{n^{c/v}_k-n^{c/v}_{-k}}{\tau_k}.\label{Gnetaphn}
\end{gather}

\emph{Intraband energy relaxation.}
This intraband energy relaxation mechanism couples electrons to an external bath of bosons at temperature $T$ (for example phonons), enabling the exchange of a small quantity of energy $\delta E$ between electrons at a rate $\Gamma_R$.

We consider a state with momentum $k$ and energy $E_k$ scattering with states at momentum $\pm(k+q_1)$ and $\pm(k-q_2)$ that satisfy energy conservation: $E_{k+q_{1/2}}=E_k\pm\delta E$.
\begin{equation*}
\begin{split}
\Gamma_E^c=\Gamma_R[&-(1+b)n_k(1-n^e_{k-q_2})+bn^e_{k-q_2}(1-n_k)- \\
     &-bn_k(1-n^e_{k+q_1})+(1+b)n^e_{k+q_1}(1-n_k)]=\\
     =\Gamma_R[&n^e_{k+q_1}-n_k+n_k(n^e_{k-q_2}-n^e_{k+q_1})+ \\
     &+b(n^e_{k-q_2}+n^e_{k+q_1}-2n_k)].
\end{split}
\end{equation*}

We calculate the even and odd part of the scattering rate, expanding for small $q_{1/2}$ and using $b\approx T/\delta E$:
\begin{gather}
\label{GRe}\Gamma_{E,e}\approx\Gamma_R\delta E[\frac1{v_k}\partial_kn_k^e(1-2n_k^e)+\frac T{v}\partial_k\frac1{v}\partial_kn_k^e];\\
\label{GRo}\Gamma_{E,o}=-\Gamma_Rn_k^o[1+2\frac{T}{\Delta E}+2\frac{\delta E}{v}\partial_kn_k^e],
\end{gather}
where $\frac{\partial_k}v=\partial_E$ and from Eq. \eqref{gEe} $\gamma_R\equiv\Gamma_R\delta E\frac{\tau}2$.

\emph{Interband photon-assisted scattering.}
The typical energy carried by a photon is of order $t\sim1\,\rm{eV}$, corresponding to a wave vector $q\sim t/(\hbar c)\sim5\cdot10^{-8}\um{{\mbox{\AA}}^{-1}}$ $\ll1/a\sim0.2\um{{\mbox{\AA}}^{-1}}$; thus the photon momentum is negligible compared to the typical electron momentum and the transition can be considered vertical.

The matrix element is $\mathcal A^2_k=\Gamma_0\Delta^2\cos^2(ka)/E_k$ with $\Gamma_0\sim10^9\um{s^{-1}}$; the change rate for electrons in conduction band is given by the difference between up scattering and down scattering:
\begin{gather}
\Gamma_{I,\textrm{up}}(k)=\mathcal A^2_k(1-n^c_k)n^v_kb;\notag\\
\Gamma_{I,\textrm{down}}(k)=\mathcal A^2_kn^c_k(1-n^v_k)(1+b);\label{Gnetpht}\\
\Gamma_I=\Gamma_{I,\textrm{down}}-\Gamma_{I,\textrm{up}}.\notag
\end{gather}

\emph{Interband optical-phonon-assisted scattering.}
The optical phonons have a constant dispersion relation $\omega_{\textrm{opt}}(k)=\omega_{\textrm{ph}}\sim0.1$. This interband scattering occurs for $\omega_{\textrm{ph}}>2\dl$ and is qualitatively similar to that mediated by photons, but the transitions are not exactly vertical. Nevertheless, $\omega$ is small and the scattering is limited to a tiny region around the gap ($|k|\lesssim\omega_{\textrm{ph}}/2$); we neglect variations of $n_k$ within this region and treat the scattering as vertical. Analogously we have up and down scattering with rate $\Gamma_0^{opt}\sim10^{10}\um{s^{-1}}$:
\begin{gather}
\Gamma^{\textrm{phn}}_{I,\textrm{up}}(k)=(\mathcal A^{\textrm{phn}}_k)^2(1-n^c_k)n^v_kb; \notag\\
\Gamma^{\textrm{phn}}_{I,\textrm{down}}(k)=(\mathcal A^{\textrm{phn}}_k)^2n^c_k(1-n^v_k)(1+b); \label{Gnetophn}\\
\Gamma^{\textrm{phn}}_I=\Gamma^{\textrm{phn}}_{I,\textrm{down}}-\Gamma^{\textrm{phn}}_{I,\textrm{up}}; \notag \\
(\mathcal A^{\textrm{phn}}_k)^2=\Gamma_0^{opt}\frac{[(\omega_{\textrm{ph}}-\epk)^2+\epk^2-2\dl^2](\omega_{\textrm{ph}}-\epk)}{8\sqrt{(\omega_{\textrm{ph}}-\epk)^2-\dl^2}}.\label{Akphn}
\end{gather}

\emph{Landau-Zener tunneling.} The electric field creates a non-zero probability of tunneling between the two bands:
\begin{gather}\label{LZ1}
\Gamma_Z^{(0)}=\frac{eFa}{\pi\hbar}e^{-\pi\dl^2/2teFa};\\
\label{LZg0}\gamma_{Z,0}=\frac{\El}{2\pi}e^{-\frac{\pi\dl^2}{2teFa}}\frac{e^{-\frac{\pi\varepsilon_k^2}{2teFa}}}{\sqrt{2t/eFa}};\\
\label{LZg}\gamma_Z=\gamma_{Z,0}(n^v_k-n^c_k).
\end{gather}
In Eq. \eqref{LZ1} $\Gamma_Z^{(0)}$ is the total scattering rate \cite{ZL:Landau, Zener:two}. In calculating $\gamma_{Z,0}$ in Eq. \eqref{LZg0}, we observed that the tunneling occurs preferentially near the gap and modeled this behavior by assuming a gaussian dependence on the energy $E_k^2=\varepsilon_k^2+\dl^2$ and normalizing the $k$-dependent part.

\section{Boltzmann equation solution for $\dl>T$}

In this appendix we study the Boltzmann equation Eq. \eqref{finaleq}. We integrate Eq. \eqref{nogb} directly, getting
\begin{subequations}
\begin{gather}
\partial_En_e=\frac{C}{T_{\textrm{eff}}}e^{-\phi(E)}; \label{nogbdn}\\
\phi(E)=\int_\Delta^{E}dE'
\frac{\gamma_R+\mathcal{E}^2v\partial_{E^\prime}v}{\gamma_RT+\mathcal{E}^2v^2}, \label{phidef}
\end{gather}
\end{subequations}
where $C$ is an arbitrary constant to be determined by requiring that the interband up and down scattering processes balance on average.

Integration of Eq. \eqref{nogbdn} determines $n_e$ except for an additive constant that we determine assuming that the range of energies is infinite and requiring $n\rightarrow 0$ as $E\rightarrow\infty$; this is equivalent to have a vanishing net particle current at infinite energy, which is the most physical condition. The result is
\begin{equation}\label{solutionnogb}
n_e=\frac{C}{T_{\textrm{eff}}}\int_E^\infty dE' e^{-\phi(E')}.
\end{equation}

We now introduce an energy cutoff $E_c\gg\Delta$ in order to deal with the finite range of energies of the Brillouin zone. For $E>E_c$, $\phi(E)\approx \frac{E}{T_{\textrm{eff}}}$ and the distribution is exponentially decreasing, while for $E<E_c$ we write
\begin{equation}\label{decompose}
n_e=\frac{C}{T_{\textrm{eff}}}\left(\int_{E_c}^\infty +\int_E^{E_c}\right)dE' e^{-\phi(E^\prime)}.
\end{equation}

We rearrange the first term and obtain
\begin{gather}
n_e=Ce^{-\phi(E_c)}
+\frac{C}{T_{\textrm{eff}}}\int_E^{E_c} dE^\prime e^{-\phi(E^\prime)},\,\,\,\,E<E_c;\label{solutionnogb1}\\
\label{solnogb2}n_e\approx Ce^{-(E-E_c)/T_{\textrm{eff}}}e^{-\phi(E_c)},\,\,\,\ E\geq E_c.
\end{gather}

We may evaluate $\phi$ analytically: we use the $E<E_c$ expression for the velocity $v=v_F\sqrt{E^2-\dl^2}/E$ and write
\begin{equation}
\phi(E)=\int_\Delta^EdE^\prime\frac{1}{T_{\textrm{eff}}}\frac{(E^\prime)^2+\frac{T_{\textrm{eff}}-T}{E^\prime}\Delta^2}{(E^\prime)^2-\frac{T_{\textrm{eff}}-T}{T_{\textrm{eff}}}\Delta^2}.
\label{beta}
\end{equation}

We recognize that the integrand in $\phi(E)$ can be written as $1/T_{eff}$ plus an energy dependent contribution which can be integrated using Eq. \eqref{phidef}, leading to
\begin{equation}\label{phiE}
\phi(E)=\frac{E-\Delta}{T_{\textrm{eff}}}+\frac12\sum_{\pm}\left(1\mp\frac{\Theta\dl}{T_{\textrm{eff}}}\right)\ln\left(\frac{1\pm\frac{\Theta\Delta}{E}}{1\pm\Theta}\right),
\end{equation}
where $\Theta=\sqrt{(T_{\textrm{eff}}-T)/T_{\textrm{eff}}}$. The combination of Eq. \eqref{solutionnogb1}, \eqref{solnogb2} and \eqref{phiE} allows to evaluate $n_e$ for all energies and then determine $C$.

\bibliographystyle{apsrev4-1}
\bibliography{Bibliography}{}

\begin{thebibliography}{25}%
\makeatletter
\providecommand \@ifxundefined [1]{%
 \@ifx{#1\undefined}
}%
\providecommand \@ifnum [1]{%
 \ifnum #1\expandafter \@firstoftwo
 \else \expandafter \@secondoftwo
 \fi
}%
\providecommand \@ifx [1]{%
 \ifx #1\expandafter \@firstoftwo
 \else \expandafter \@secondoftwo
 \fi
}%
\providecommand \natexlab [1]{#1}%
\providecommand \enquote  [1]{``#1''}%
\providecommand \bibnamefont  [1]{#1}%
\providecommand \bibfnamefont [1]{#1}%
\providecommand \citenamefont [1]{#1}%
\providecommand \href@noop [0]{\@secondoftwo}%
\providecommand \href [0]{\begingroup \@sanitize@url \@href}%
\providecommand \@href[1]{\@@startlink{#1}\@@href}%
\providecommand \@@href[1]{\endgroup#1\@@endlink}%
\providecommand \@sanitize@url [0]{\catcode `\\12\catcode `\$12\catcode
  `\&12\catcode `\#12\catcode `\^12\catcode `\_12\catcode `\%12\relax}%
\providecommand \@@startlink[1]{}%
\providecommand \@@endlink[0]{}%
\providecommand \url  [0]{\begingroup\@sanitize@url \@url }%
\providecommand \@url [1]{\endgroup\@href {#1}{\urlprefix }}%
\providecommand \urlprefix  [0]{URL }%
\providecommand \Eprint [0]{\href }%
\providecommand \doibase [0]{http://dx.doi.org/}%
\providecommand \selectlanguage [0]{\@gobble}%
\providecommand \bibinfo  [0]{\@secondoftwo}%
\providecommand \bibfield  [0]{\@secondoftwo}%
\providecommand \translation [1]{[#1]}%
\providecommand \BibitemOpen [0]{}%
\providecommand \bibitemStop [0]{}%
\providecommand \bibitemNoStop [0]{.\EOS\space}%
\providecommand \EOS [0]{\spacefactor3000\relax}%
\providecommand \BibitemShut  [1]{\csname bibitem#1\endcsname}%
\let\auto@bib@innerbib\@empty
%</preamble>
\bibitem [{\citenamefont {Imada}\ \emph {et~al.}(1998)\citenamefont {Imada},
  \citenamefont {Fujimori},\ and\ \citenamefont {Tokura}}]{MIT:review}%
  \BibitemOpen
  \bibfield  {author} {\bibinfo {author} {\bibfnamefont {M.}~\bibnamefont
  {Imada}}, \bibinfo {author} {\bibfnamefont {A.}~\bibnamefont {Fujimori}}, \
  and\ \bibinfo {author} {\bibfnamefont {Y.}~\bibnamefont {Tokura}},\ }\href
  {\doibase 10.1103/RevModPhys.70.1039} {\bibfield  {journal} {\bibinfo
  {journal} {Rev. Mod. Phys.}\ }\textbf {\bibinfo {volume} {70}},\ \bibinfo
  {pages} {1039} (\bibinfo {year} {1998})}\BibitemShut {NoStop}%
\bibitem [{\citenamefont {Averitt}\ \emph
  {et~al.}(2001{\natexlab{a}})\citenamefont {Averitt}, \citenamefont
  {Rodriguez}, \citenamefont {Lobad}, \citenamefont {Siders}, \citenamefont
  {Trugman},\ and\ \citenamefont {Taylor}}]{Aver:optCu}%
  \BibitemOpen
  \bibfield  {author} {\bibinfo {author} {\bibfnamefont {R.~D.}\ \bibnamefont
  {Averitt}}, \bibinfo {author} {\bibfnamefont {G.}~\bibnamefont {Rodriguez}},
  \bibinfo {author} {\bibfnamefont {A.~I.}\ \bibnamefont {Lobad}}, \bibinfo
  {author} {\bibfnamefont {J.~L.~W.}\ \bibnamefont {Siders}}, \bibinfo {author}
  {\bibfnamefont {S.~A.}\ \bibnamefont {Trugman}}, \ and\ \bibinfo {author}
  {\bibfnamefont {A.~J.}\ \bibnamefont {Taylor}},\ }\href {\doibase
  10.1103/PhysRevB.63.140502} {\bibfield  {journal} {\bibinfo  {journal} {Phys.
  Rev. B}\ }\textbf {\bibinfo {volume} {63}},\ \bibinfo {pages} {140502}
  (\bibinfo {year} {2001}{\natexlab{a}})}\BibitemShut {NoStop}%
\bibitem [{\citenamefont {Averitt}\ \emph
  {et~al.}(2001{\natexlab{b}})\citenamefont {Averitt}, \citenamefont {Lobad},
  \citenamefont {Kwon}, \citenamefont {Trugman}, \citenamefont
  {Thorsm\o{}lle},\ and\ \citenamefont {Taylor}}]{Aver:optLa}%
  \BibitemOpen
  \bibfield  {author} {\bibinfo {author} {\bibfnamefont {R.~D.}\ \bibnamefont
  {Averitt}}, \bibinfo {author} {\bibfnamefont {A.~I.}\ \bibnamefont {Lobad}},
  \bibinfo {author} {\bibfnamefont {C.}~\bibnamefont {Kwon}}, \bibinfo {author}
  {\bibfnamefont {S.~A.}\ \bibnamefont {Trugman}}, \bibinfo {author}
  {\bibfnamefont {V.~K.}\ \bibnamefont {Thorsm\o{}lle}}, \ and\ \bibinfo
  {author} {\bibfnamefont {A.~J.}\ \bibnamefont {Taylor}},\ }\href {\doibase
  10.1103/PhysRevLett.87.017401} {\bibfield  {journal} {\bibinfo  {journal}
  {Phys. Rev. Lett.}\ }\textbf {\bibinfo {volume} {87}},\ \bibinfo {pages}
  {017401} (\bibinfo {year} {2001}{\natexlab{b}})}\BibitemShut {NoStop}%
\bibitem [{\citenamefont {Averitt}\ and\ \citenamefont
  {Taylor}(2002)}]{Aver:opt-rev}%
  \BibitemOpen
  \bibfield  {author} {\bibinfo {author} {\bibfnamefont {R.~D.}\ \bibnamefont
  {Averitt}}\ and\ \bibinfo {author} {\bibfnamefont {A.~J.}\ \bibnamefont
  {Taylor}},\ }\href {http://stacks.iop.org/0953-8984/14/i=50/a=203} {\bibfield
   {journal} {\bibinfo  {journal} {Journal of Physics: Condensed Matter}\
  }\textbf {\bibinfo {volume} {14}},\ \bibinfo {pages} {R1357} (\bibinfo {year}
  {2002})}\BibitemShut {NoStop}%
\bibitem [{\citenamefont {Winkler}\ \emph {et~al.}(2011)\citenamefont
  {Winkler}, \citenamefont {Recht}, \citenamefont {Sher}, \citenamefont {Said},
  \citenamefont {Mazur},\ and\ \citenamefont {Aziz}}]{Mazur}%
  \BibitemOpen
  \bibfield  {author} {\bibinfo {author} {\bibfnamefont {M.~T.}\ \bibnamefont
  {Winkler}}, \bibinfo {author} {\bibfnamefont {D.}~\bibnamefont {Recht}},
  \bibinfo {author} {\bibfnamefont {M.-J.}\ \bibnamefont {Sher}}, \bibinfo
  {author} {\bibfnamefont {A.~J.}\ \bibnamefont {Said}}, \bibinfo {author}
  {\bibfnamefont {E.}~\bibnamefont {Mazur}}, \ and\ \bibinfo {author}
  {\bibfnamefont {M.~J.}\ \bibnamefont {Aziz}},\ }\href {\doibase
  10.1103/PhysRevLett.106.178701} {\bibfield  {journal} {\bibinfo  {journal}
  {Phys. Rev. Lett.}\ }\textbf {\bibinfo {volume} {106}},\ \bibinfo {pages}
  {178701} (\bibinfo {year} {2011})}\BibitemShut {NoStop}%
\bibitem [{\citenamefont {Ao}\ \emph {et~al.}(2006)\citenamefont {Ao},
  \citenamefont {Ping}, \citenamefont {Widmann}, \citenamefont {Price},
  \citenamefont {Lee}, \citenamefont {Tam}, \citenamefont {Springer},\ and\
  \citenamefont {Ng}}]{laser:gold}%
  \BibitemOpen
  \bibfield  {author} {\bibinfo {author} {\bibfnamefont {T.}~\bibnamefont
  {Ao}}, \bibinfo {author} {\bibfnamefont {Y.}~\bibnamefont {Ping}}, \bibinfo
  {author} {\bibfnamefont {K.}~\bibnamefont {Widmann}}, \bibinfo {author}
  {\bibfnamefont {D.~F.}\ \bibnamefont {Price}}, \bibinfo {author}
  {\bibfnamefont {E.}~\bibnamefont {Lee}}, \bibinfo {author} {\bibfnamefont
  {H.}~\bibnamefont {Tam}}, \bibinfo {author} {\bibfnamefont {P.~T.}\
  \bibnamefont {Springer}}, \ and\ \bibinfo {author} {\bibfnamefont
  {A.}~\bibnamefont {Ng}},\ }\href {\doibase 10.1103/PhysRevLett.96.055001}
  {\bibfield  {journal} {\bibinfo  {journal} {Phys. Rev. Lett.}\ }\textbf
  {\bibinfo {volume} {96}},\ \bibinfo {pages} {055001} (\bibinfo {year}
  {2006})}\BibitemShut {NoStop}%
\bibitem [{\citenamefont {Basov}\ \emph {et~al.}(2011)\citenamefont {Basov},
  \citenamefont {Averitt}, \citenamefont {van~der Marel}, \citenamefont
  {Dressel},\ and\ \citenamefont {Haule}}]{Basov:review}%
  \BibitemOpen
  \bibfield  {author} {\bibinfo {author} {\bibfnamefont {D.~N.}\ \bibnamefont
  {Basov}}, \bibinfo {author} {\bibfnamefont {R.~D.}\ \bibnamefont {Averitt}},
  \bibinfo {author} {\bibfnamefont {D.}~\bibnamefont {van~der Marel}}, \bibinfo
  {author} {\bibfnamefont {M.}~\bibnamefont {Dressel}}, \ and\ \bibinfo
  {author} {\bibfnamefont {K.}~\bibnamefont {Haule}},\ }\href {\doibase
  10.1103/RevModPhys.83.471} {\bibfield  {journal} {\bibinfo  {journal} {Rev.
  Mod. Phys.}\ }\textbf {\bibinfo {volume} {83}},\ \bibinfo {pages} {471}
  (\bibinfo {year} {2011})}\BibitemShut {NoStop}%
\bibitem [{\citenamefont {Mitrano}\ \emph {et~al.}(2016)\citenamefont
  {Mitrano}, \citenamefont {Cantaluppi}, \citenamefont {Nicoletti},
  \citenamefont {Kaiser}, \citenamefont {Perucchi}, \citenamefont {Lupi},
  \citenamefont {Pietro}, \citenamefont {Pontiroli}, \citenamefont {Riccò},
  \citenamefont {Clark}, \citenamefont {Jaksch},\ and\ \citenamefont
  {Cavalleri}}]{Cavalleri:K3C60}%
  \BibitemOpen
  \bibfield  {author} {\bibinfo {author} {\bibfnamefont {M.}~\bibnamefont
  {Mitrano}}, \bibinfo {author} {\bibfnamefont {A.}~\bibnamefont {Cantaluppi}},
  \bibinfo {author} {\bibfnamefont {D.}~\bibnamefont {Nicoletti}}, \bibinfo
  {author} {\bibfnamefont {S.}~\bibnamefont {Kaiser}}, \bibinfo {author}
  {\bibfnamefont {A.}~\bibnamefont {Perucchi}}, \bibinfo {author}
  {\bibfnamefont {S.}~\bibnamefont {Lupi}}, \bibinfo {author} {\bibfnamefont
  {P.~D.}\ \bibnamefont {Pietro}}, \bibinfo {author} {\bibfnamefont
  {D.}~\bibnamefont {Pontiroli}}, \bibinfo {author} {\bibfnamefont
  {M.}~\bibnamefont {Riccò}}, \bibinfo {author} {\bibfnamefont {S.~R.}\
  \bibnamefont {Clark}}, \bibinfo {author} {\bibfnamefont {D.}~\bibnamefont
  {Jaksch}}, \ and\ \bibinfo {author} {\bibfnamefont {A.}~\bibnamefont
  {Cavalleri}},\ }\href@noop {} {\bibfield  {journal} {\bibinfo  {journal}
  {Nature}\ }\textbf {\bibinfo {volume} {530}},\ \bibinfo {pages} {461–464}
  (\bibinfo {year} {2016})}\BibitemShut {NoStop}%
\bibitem [{\citenamefont {Averitt}\ \emph {et~al.}(2002)\citenamefont
  {Averitt}, \citenamefont {Thorsmølle}, \citenamefont {Jia}, \citenamefont
  {Trugman},\ and\ \citenamefont {Taylor}}]{Aver:THz}%
  \BibitemOpen
  \bibfield  {author} {\bibinfo {author} {\bibfnamefont {R.}~\bibnamefont
  {Averitt}}, \bibinfo {author} {\bibfnamefont {V.}~\bibnamefont
  {Thorsmølle}}, \bibinfo {author} {\bibfnamefont {Q.}~\bibnamefont {Jia}},
  \bibinfo {author} {\bibfnamefont {S.}~\bibnamefont {Trugman}}, \ and\
  \bibinfo {author} {\bibfnamefont {A.}~\bibnamefont {Taylor}},\ }\href
  {\doibase https://doi.org/10.1016/S0921-4526(01)01166-8} {\bibfield
  {journal} {\bibinfo  {journal} {Physica B: Condensed Matter}\ }\textbf
  {\bibinfo {volume} {312-313}},\ \bibinfo {pages} {86 } (\bibinfo {year}
  {2002})}\BibitemShut {NoStop}%
\bibitem [{\citenamefont {Mitra}\ \emph {et~al.}(2006)\citenamefont {Mitra},
  \citenamefont {Takei}, \citenamefont {Kim},\ and\ \citenamefont
  {Millis}}]{Mitra06}%
  \BibitemOpen
  \bibfield  {author} {\bibinfo {author} {\bibfnamefont {A.}~\bibnamefont
  {Mitra}}, \bibinfo {author} {\bibfnamefont {S.}~\bibnamefont {Takei}},
  \bibinfo {author} {\bibfnamefont {Y.~B.}\ \bibnamefont {Kim}}, \ and\
  \bibinfo {author} {\bibfnamefont {A.~J.}\ \bibnamefont {Millis}},\ }\href
  {\doibase 10.1103/PhysRevLett.97.236808} {\bibfield  {journal} {\bibinfo
  {journal} {Phys. Rev. Lett.}\ }\textbf {\bibinfo {volume} {97}},\ \bibinfo
  {pages} {236808} (\bibinfo {year} {2006})}\BibitemShut {NoStop}%
\bibitem [{\citenamefont {Kanki}\ \emph {et~al.}(2012)\citenamefont {Kanki},
  \citenamefont {Kawatani}, \citenamefont {Takami},\ and\ \citenamefont
  {Tanaka}}]{Maecit:7}%
  \BibitemOpen
  \bibfield  {author} {\bibinfo {author} {\bibfnamefont {T.}~\bibnamefont
  {Kanki}}, \bibinfo {author} {\bibfnamefont {K.}~\bibnamefont {Kawatani}},
  \bibinfo {author} {\bibfnamefont {H.}~\bibnamefont {Takami}}, \ and\ \bibinfo
  {author} {\bibfnamefont {H.}~\bibnamefont {Tanaka}},\ }\href@noop {}
  {\bibfield  {journal} {\bibinfo  {journal} {Applied Physics Letters}\
  }\textbf {\bibinfo {volume} {101}},\ \bibinfo {pages} {243118} (\bibinfo
  {year} {2012})}\BibitemShut {NoStop}%
\bibitem [{\citenamefont {Iwasa}\ \emph {et~al.}(1989)\citenamefont {Iwasa},
  \citenamefont {Koda}, \citenamefont {Tokura}, \citenamefont {Koshihara},
  \citenamefont {Iwasawa},\ and\ \citenamefont {Saito}}]{Maecit:8}%
  \BibitemOpen
  \bibfield  {author} {\bibinfo {author} {\bibfnamefont {Y.}~\bibnamefont
  {Iwasa}}, \bibinfo {author} {\bibfnamefont {T.}~\bibnamefont {Koda}},
  \bibinfo {author} {\bibfnamefont {Y.}~\bibnamefont {Tokura}}, \bibinfo
  {author} {\bibfnamefont {S.}~\bibnamefont {Koshihara}}, \bibinfo {author}
  {\bibfnamefont {N.}~\bibnamefont {Iwasawa}}, \ and\ \bibinfo {author}
  {\bibfnamefont {G.}~\bibnamefont {Saito}},\ }\href@noop {} {\bibfield
  {journal} {\bibinfo  {journal} {Applied physics letters}\ }\textbf {\bibinfo
  {volume} {55}},\ \bibinfo {pages} {2111} (\bibinfo {year}
  {1989})}\BibitemShut {NoStop}%
\bibitem [{\citenamefont {Yamanouchi}\ \emph {et~al.}(1999)\citenamefont
  {Yamanouchi}, \citenamefont {Taguchi},\ and\ \citenamefont
  {Tokura}}]{Maecit:9}%
  \BibitemOpen
  \bibfield  {author} {\bibinfo {author} {\bibfnamefont {S.}~\bibnamefont
  {Yamanouchi}}, \bibinfo {author} {\bibfnamefont {Y.}~\bibnamefont {Taguchi}},
  \ and\ \bibinfo {author} {\bibfnamefont {Y.}~\bibnamefont {Tokura}},\
  }\href@noop {} {\bibfield  {journal} {\bibinfo  {journal} {Physical Review
  Letters}\ }\textbf {\bibinfo {volume} {83}},\ \bibinfo {pages} {5555}
  (\bibinfo {year} {1999})}\BibitemShut {NoStop}%
\bibitem [{\citenamefont {Taguchi}\ \emph {et~al.}(2000)\citenamefont
  {Taguchi}, \citenamefont {Matsumoto},\ and\ \citenamefont
  {Tokura}}]{Maecit:10}%
  \BibitemOpen
  \bibfield  {author} {\bibinfo {author} {\bibfnamefont {Y.}~\bibnamefont
  {Taguchi}}, \bibinfo {author} {\bibfnamefont {T.}~\bibnamefont {Matsumoto}},
  \ and\ \bibinfo {author} {\bibfnamefont {Y.}~\bibnamefont {Tokura}},\
  }\href@noop {} {\bibfield  {journal} {\bibinfo  {journal} {Physical Review
  B}\ }\textbf {\bibinfo {volume} {62}},\ \bibinfo {pages} {7015} (\bibinfo
  {year} {2000})}\BibitemShut {NoStop}%
\bibitem [{\citenamefont {Hatsuda}\ \emph {et~al.}(2003)\citenamefont
  {Hatsuda}, \citenamefont {Kimura},\ and\ \citenamefont {Tokura}}]{Maecit:11}%
  \BibitemOpen
  \bibfield  {author} {\bibinfo {author} {\bibfnamefont {K.}~\bibnamefont
  {Hatsuda}}, \bibinfo {author} {\bibfnamefont {T.}~\bibnamefont {Kimura}}, \
  and\ \bibinfo {author} {\bibfnamefont {Y.}~\bibnamefont {Tokura}},\
  }\href@noop {} {\bibfield  {journal} {\bibinfo  {journal} {Applied physics
  letters}\ }\textbf {\bibinfo {volume} {83}},\ \bibinfo {pages} {3329}
  (\bibinfo {year} {2003})}\BibitemShut {NoStop}%
\bibitem [{\citenamefont {Guiot}\ \emph {et~al.}(2013)\citenamefont {Guiot},
  \citenamefont {Cario}, \citenamefont {Janod}, \citenamefont {Corraze},
  \citenamefont {Ta~Phuoc}, \citenamefont {Rozenberg}, \citenamefont {Stoliar},
  \citenamefont {Cren},\ and\ \citenamefont {Roditchev}}]{GaTaSe}%
  \BibitemOpen
  \bibfield  {author} {\bibinfo {author} {\bibfnamefont {V.}~\bibnamefont
  {Guiot}}, \bibinfo {author} {\bibfnamefont {L.}~\bibnamefont {Cario}},
  \bibinfo {author} {\bibfnamefont {E.}~\bibnamefont {Janod}}, \bibinfo
  {author} {\bibfnamefont {B.}~\bibnamefont {Corraze}}, \bibinfo {author}
  {\bibfnamefont {V.}~\bibnamefont {Ta~Phuoc}}, \bibinfo {author}
  {\bibfnamefont {M.}~\bibnamefont {Rozenberg}}, \bibinfo {author}
  {\bibfnamefont {P.}~\bibnamefont {Stoliar}}, \bibinfo {author} {\bibfnamefont
  {T.}~\bibnamefont {Cren}}, \ and\ \bibinfo {author} {\bibfnamefont
  {D.}~\bibnamefont {Roditchev}},\ }\href@noop {} {\bibfield  {journal}
  {\bibinfo  {journal} {Nature Communications}\ }\textbf {\bibinfo {volume}
  {4}},\ \bibinfo {pages} {1722} (\bibinfo {year} {2013})}\BibitemShut
  {NoStop}%
\bibitem [{\citenamefont {Nakamura}\ \emph {et~al.}(2013)\citenamefont
  {Nakamura}, \citenamefont {Sakaki}, \citenamefont {Yamanaka}, \citenamefont
  {Tamaru}, \citenamefont {Suzuki},\ and\ \citenamefont {Maeno}}]{Maeno:mainp}%
  \BibitemOpen
  \bibfield  {author} {\bibinfo {author} {\bibfnamefont {F.}~\bibnamefont
  {Nakamura}}, \bibinfo {author} {\bibfnamefont {M.}~\bibnamefont {Sakaki}},
  \bibinfo {author} {\bibfnamefont {Y.}~\bibnamefont {Yamanaka}}, \bibinfo
  {author} {\bibfnamefont {S.}~\bibnamefont {Tamaru}}, \bibinfo {author}
  {\bibfnamefont {T.}~\bibnamefont {Suzuki}}, \ and\ \bibinfo {author}
  {\bibfnamefont {Y.}~\bibnamefont {Maeno}},\ }\href@noop {} {\bibfield
  {journal} {\bibinfo  {journal} {Scientific reports}\ }\textbf {\bibinfo
  {volume} {3}},\ \bibinfo {pages} {2536} (\bibinfo {year} {2013})}\BibitemShut
  {NoStop}%
\bibitem [{\citenamefont {Okazaki}\ \emph {et~al.}(2013)\citenamefont
  {Okazaki}, \citenamefont {Nishina}, \citenamefont {Yasui}, \citenamefont
  {Nakamura}, \citenamefont {Suzuki},\ and\ \citenamefont
  {Terasaki}}]{Fuku:CRO}%
  \BibitemOpen
  \bibfield  {author} {\bibinfo {author} {\bibfnamefont {R.}~\bibnamefont
  {Okazaki}}, \bibinfo {author} {\bibfnamefont {Y.}~\bibnamefont {Nishina}},
  \bibinfo {author} {\bibfnamefont {Y.}~\bibnamefont {Yasui}}, \bibinfo
  {author} {\bibfnamefont {F.}~\bibnamefont {Nakamura}}, \bibinfo {author}
  {\bibfnamefont {T.}~\bibnamefont {Suzuki}}, \ and\ \bibinfo {author}
  {\bibfnamefont {I.}~\bibnamefont {Terasaki}},\ }\href@noop {} {\bibfield
  {journal} {\bibinfo  {journal} {Journal of the Physical Society of Japan}\
  }\textbf {\bibinfo {volume} {82}},\ \bibinfo {pages} {103702} (\bibinfo
  {year} {2013})}\BibitemShut {NoStop}%
\bibitem [{\citenamefont {Gorelov}\ \emph {et~al.}(2010)\citenamefont
  {Gorelov}, \citenamefont {Karolak}, \citenamefont {Wehling}, \citenamefont
  {Lechermann}, \citenamefont {Lichtenstein},\ and\ \citenamefont
  {Pavarini}}]{Pavarini:CRO}%
  \BibitemOpen
  \bibfield  {author} {\bibinfo {author} {\bibfnamefont {E.}~\bibnamefont
  {Gorelov}}, \bibinfo {author} {\bibfnamefont {M.}~\bibnamefont {Karolak}},
  \bibinfo {author} {\bibfnamefont {T.~O.}\ \bibnamefont {Wehling}}, \bibinfo
  {author} {\bibfnamefont {F.}~\bibnamefont {Lechermann}}, \bibinfo {author}
  {\bibfnamefont {A.~I.}\ \bibnamefont {Lichtenstein}}, \ and\ \bibinfo
  {author} {\bibfnamefont {E.}~\bibnamefont {Pavarini}},\ }\href@noop {}
  {\bibfield  {journal} {\bibinfo  {journal} {Physical review letters}\
  }\textbf {\bibinfo {volume} {104}},\ \bibinfo {pages} {226401} (\bibinfo
  {year} {2010})}\BibitemShut {NoStop}%
\bibitem [{\citenamefont {Friedt}\ \emph {et~al.}(2001)\citenamefont {Friedt},
  \citenamefont {Braden}, \citenamefont {Andr{\'e}}, \citenamefont {Adelmann},
  \citenamefont {Nakatsuji},\ and\ \citenamefont {Maeno}}]{Friedt:CROPT}%
  \BibitemOpen
  \bibfield  {author} {\bibinfo {author} {\bibfnamefont {O.}~\bibnamefont
  {Friedt}}, \bibinfo {author} {\bibfnamefont {M.}~\bibnamefont {Braden}},
  \bibinfo {author} {\bibfnamefont {G.}~\bibnamefont {Andr{\'e}}}, \bibinfo
  {author} {\bibfnamefont {P.}~\bibnamefont {Adelmann}}, \bibinfo {author}
  {\bibfnamefont {S.}~\bibnamefont {Nakatsuji}}, \ and\ \bibinfo {author}
  {\bibfnamefont {Y.}~\bibnamefont {Maeno}},\ }\href@noop {} {\bibfield
  {journal} {\bibinfo  {journal} {Physical Review B}\ }\textbf {\bibinfo
  {volume} {63}},\ \bibinfo {pages} {174432} (\bibinfo {year}
  {2001})}\BibitemShut {NoStop}%
\bibitem [{\citenamefont {Amaricci}\ \emph {et~al.}(2012)\citenamefont
  {Amaricci}, \citenamefont {Weber}, \citenamefont {Capone},\ and\
  \citenamefont {Kotliar}}]{Amaricci12}%
  \BibitemOpen
  \bibfield  {author} {\bibinfo {author} {\bibfnamefont {A.}~\bibnamefont
  {Amaricci}}, \bibinfo {author} {\bibfnamefont {C.}~\bibnamefont {Weber}},
  \bibinfo {author} {\bibfnamefont {M.}~\bibnamefont {Capone}}, \ and\ \bibinfo
  {author} {\bibfnamefont {G.}~\bibnamefont {Kotliar}},\ }\href {\doibase
  10.1103/PhysRevB.86.085110} {\bibfield  {journal} {\bibinfo  {journal} {Phys.
  Rev. B}\ }\textbf {\bibinfo {volume} {86}},\ \bibinfo {pages} {085110}
  (\bibinfo {year} {2012})}\BibitemShut {NoStop}%
\bibitem [{\citenamefont {Landau}(1932)}]{ZL:Landau}%
  \BibitemOpen
  \bibfield  {author} {\bibinfo {author} {\bibfnamefont {L.}~\bibnamefont
  {Landau}},\ }\href@noop {} {\bibfield  {journal} {\bibinfo  {journal}
  {Physikalische Zeitschrift der Sowjetunion}\ }\textbf {\bibinfo {volume}
  {2}},\ \bibinfo {pages} {46–51} (\bibinfo {year} {1932})}\BibitemShut
  {NoStop}%
\bibitem [{\citenamefont {Zener}(1934)}]{Zener:two}%
  \BibitemOpen
  \bibfield  {author} {\bibinfo {author} {\bibfnamefont {C.}~\bibnamefont
  {Zener}},\ }\href {http://www.jstor.org/stable/2935519} {\bibfield  {journal}
  {\bibinfo  {journal} {Proc. R. Soc. London, Ser. A}\ }\textbf {\bibinfo
  {volume} {145}},\ \bibinfo {pages} {523} (\bibinfo {year}
  {1934})}\BibitemShut {NoStop}%
\bibitem [{\citenamefont {Peierls}(1955)}]{CDW:Peierls}%
  \BibitemOpen
  \bibfield  {author} {\bibinfo {author} {\bibfnamefont {R.}~\bibnamefont
  {Peierls}},\ }\href@noop {} {\emph {\bibinfo {title} {Quantum Theory of
  Solids}}}\ (\bibinfo  {publisher} {Oxford: Clarendon},\ \bibinfo {year}
  {1955})\ p.\ \bibinfo {pages} {108}\BibitemShut {NoStop}%
\bibitem [{\citenamefont {Han}\ \emph {et~al.}(2018)\citenamefont {Han},
  \citenamefont {Li}, \citenamefont {Aron},\ and\ \citenamefont
  {Kotliar}}]{Han}%
  \BibitemOpen
  \bibfield  {author} {\bibinfo {author} {\bibfnamefont {J.~E.}\ \bibnamefont
  {Han}}, \bibinfo {author} {\bibfnamefont {J.}~\bibnamefont {Li}}, \bibinfo
  {author} {\bibfnamefont {C.}~\bibnamefont {Aron}}, \ and\ \bibinfo {author}
  {\bibfnamefont {G.}~\bibnamefont {Kotliar}},\ }\href {\doibase
  10.1103/PhysRevB.98.035145} {\bibfield  {journal} {\bibinfo  {journal} {Phys.
  Rev. B}\ }\textbf {\bibinfo {volume} {98}},\ \bibinfo {pages} {035145}
  (\bibinfo {year} {2018})}\BibitemShut {NoStop}%
\end{thebibliography}%

\end{document}